\documentclass{article}
\title{Rest Frame System for Asymptotically Flat Spacetimes}

\author{Osvaldo M. Moreschi\thanks{Member of CONICET; Electronic mail:
    moreschi@fis.uncor.edu} \and Sergio Dain\thanks{Fellowship holder from CONICOR.}\\
  FaMAF, Ciudad Universitaria, Universidad Nacional de C\'ordoba\\
  (5000) C\'ordoba Argentina}

\date{}

\usepackage{amsmath,amssymb,amsfonts}
\newtheorem{theorem}{Theorem}[section]
\newtheorem{lemma}{Lemma}[section]

\newtheorem{corollary}{Corollary}[section]

\begin{document}

\maketitle

\begin{abstract}
The notion of center of mass for an isolated system has been 
previously encoded in the definition of the so called nice sections. 
In this article we present a generalization of the proof of existence 
of solutions to the linearized equation for nice sections, and 
formalize a local existence proof of nice sections relaxing the 
radiation condition. We report on the differentiable and non-self-crossing 
properties of this family of solutions. We also give a proof 
of the global existence of nice sections.

PACS number(s): 04.20.Ha, 04.20.Cv, 04.20.-q
\end{abstract}

\section{Introduction} \label{rfIndroduction}

There is an increasing interest in obtaining results from general 
relativity that could be applied to real systems. In particular, 
the possibility of detecting gravitational waves in the near 
future, motivates the research for the description of systems 
containing very compact objects. In modeling these systems as 
isolated ones, it is possible to distinguish at least two very 
different approaches for tackling this problem. In one of them 
the description is attempted by introducing data, with some criteria, 
for the numeric solution of the field equation. In the other 
the description is attempted by trying to reduce the infinite 
degrees of freedom of the fields to some finite set, analogously 
as is done in the particle like description in Newtonian or in 
post-Newtonian dynamics. If this second approach has any chance 
of success, it will be because it is possible to ascribe, in 
a geometric invariant way, the finite degrees of freedom to the 
physical system one wants to describe.

It is important to emphasize that the present indication of the 
existence of gravitational radiation comes from the usage of 
the balance equation in the description of the binary pulsar 
PSR 1913+16. The applicability of the present derivations of 
the balance equation to the binary pulsar system is still a matter 
of discussion. However, even in the case of having a formally 
complete derivation of this equation valid at an instant of time, 
one is still faced with the issue of its validity for long periods 
of time, or for systems involving strong gravitational fields 
and possible high velocities. The nature of the problem can be 
seen by noting that the balance equation relates time's derivatives 
of the energy of the systems with time derivatives of the quadrupole 
moments of the system. Although the total energy of the system 
could be related with the unambiguously defined Bondi momentum; 
the notion of quadrupole moment has meaning only after one has 
settled down the issue of supertranslation freedom. Our construction 
of the center of mass sections at future null infinity through 
the definition of nice sections provides a procedure to give 
a precise meaning to concepts as angular momentum and multipole 
moments.

Therefore, from the study of gravitating isolated systems in 
the vicinity of the asymptotic region, i.e., future null infinity, 
one concludes that it is essential\cite{Moreschi88'} to introduce a family of 
sections of scri, embodying the notion of generalized center 
of mass. This is the analog of having a frame of reference with 
respect to which one can study the asymptotic fields. With this 
aim in ref. \cite{Moreschi88} it was introduced the `Supercenter of mass' at 
future null infinity; which was defined in terms of the so called 
nice sections. It was indicated in this reference that the notion 
of center of mass can be attained by requiring an asymptotic 
nice section for the retarded time 
 $u\rightarrow -\infty $
 to satisfy that the angular 
momentum coincides with the intrinsic angular momentum. Then, 
if the subsequent nice sections are constructed by infinitesimal 
time translations which are parallel to the Bondi momentum, one 
obtains a one parameter family of nice sections at future null 
infinity embodying the notion of center of mass.

In this work we study some mathematical properties of nice sections. 
In Section II we review the nice section equation and some of 
its basic properties. We present a more general proof of the 
existence of solutions to the linearized equation for nice sections, 
and formalize a local existence proof of nice sections in Section 
III. The differentiable properties of nice sections are studied 
in Section IV. We report on the desired non-self-crossing property 
of this family of solutions in Section V. In Section VI we present 
a proof of global existence of nice sections. Finally, several 
complementary results are presented in three Appendices.

\section{The nice section equation} \label{rfThe nice}

We will start by recalling the nice section equation. Given a 
section S of future null infinity, one can always find a Bondi 
coordinate(see for example ref. \cite{Moreschi86}) system 
 $(u,\zeta ,\bar{\zeta } )$
 such that S is given 
by the equation 
 $u=0$
. The supermomentum at S is defined\cite{Moreschi88} by
\begin{equation} \label{rf1}
P_{lm} (S)\equiv -\frac{1}{\sqrt{4\pi } } \int\limits_{S}Y_{lm} (\zeta
,\bar{\zeta } )\;\Psi \;dS^{2};
\end{equation}
where
\begin{equation} \label{rf2}
\Psi \equiv \Psi_2 +\sigma \dot{\bar{\sigma } } +\eth^2\bar{\sigma };
\end{equation}
with $\Psi_2 $   and    $\sigma$
 being the leading order asymptotic behavior of the second 
Weyl tensor component and the Bondi shear respectively, and where 
we are using the GHP notation\cite{Geroch73} for the edth operator of the 
unit sphere. It is probably worth noting that in a Bondi system\cite{Moreschi86} 
the quantity $\Psi$ is a real function on future null infinity (see 
for example equations (3.35) and (5.20) in ref. \cite{Moreschi87}).

Given another section 
 $\tilde{S} $
 of scri, one can find another Bondi system $(\tilde{u} ,\tilde{\zeta } ,\bar{\tilde{\zeta } } )$ such that 
 $\tilde{S} $
 is given by 
 $\tilde{u} =0$
. One can further demand for that Bondi 
system to be aligned with the total (Bondi) momentum at 
 $\tilde{S} $
; i.e., 
that the 
 $l=1$
 components of the momentum are zero:
\[
\tilde{P} _{l=1,m} (\tilde{S} )=0.
\]
The relation between the new and the original Bondi system is 
given by the BMS transformation
\begin{equation} \label{rf3}
\tilde{u} =K(\zeta ,\bar{\zeta } )\left( u-\gamma (\zeta ,\bar{\zeta }
)\right),
\end{equation}

\begin{equation} \label{rf4}
\bar{\zeta } =\frac{a\;\zeta +b}{c\,\bar{\zeta } +d}
\end{equation}
with
\[
ad-bc=1,  \quad                 
K=\frac{(1+\zeta \bar{\zeta }
)}{(a\;\zeta +b)(\bar{a} \;\bar{\zeta } +\bar{b} )+(c\;\zeta +d)(\bar{c}
\;\bar{\zeta } +\bar{d} )}
;
\]
where a, b, c and d are complex constants. Note that 
 $\tilde{S} $
 can also 
be determined by 
 $u=\gamma (\zeta ,\bar{\zeta } )$
.

The section 
 $\tilde{S} $
 is said to be of the nice\cite{Moreschi88} type if
\begin{equation} \label{rf5}
\tilde{P}_{lm} (\tilde{S} )=0  \quad  \text{for} \quad       l\neq 0;
\end{equation}
that is all the `spacelike' components of the supermomentum, 
when calculated with respect to the adapted Bondi system, are 
zero; and the only nonvanishing one, namely 
 $\tilde{P} _{00} (\tilde{S} )$
, coincide with the 
total Bondi mass at 
 $\tilde{S}$.

This is the original version of the nice section equation; however 
its information is hidden in the simplicity of its form. Since 
the nice section can be determined by the supertranslation 
 $\gamma (\zeta ,\bar{\zeta } )$
, 
mentioned above, it is more useful to restate the nice section 
equation in terms of an equation for 
 $\gamma (\zeta ,\bar{\zeta } )$
.

The supermomentum at 
 $\tilde{S} $
 can also be expressed in terms of the original 
Bondi coordinate system\cite{Moreschi88}:
\begin{equation} \label{rf6}
\begin{split} 
\tilde P_{lm} (\tilde S) &\equiv -\frac{1}{4\pi} \int_{\tilde S} Y_{lm} (\tilde \zeta, \bar {\tilde \zeta} ) \tilde \Psi \, d\tilde{S}^2\\
&= -\frac{1}{4\pi} K_{lm}^{l'm'} \int_{\tilde{S}=(u=\gamma)} Y_{l'm'}(\zeta, \bar \zeta) ( \Psi(u=\gamma)- \eth^2 \bar \eth^2 \gamma) \,dS^2\\
\end{split}
\end{equation}
where the matrix $K_{lm} ^{l'm'} $
 is the transformation matrix of the generators 
of supertranslations, that is:
\begin{equation} \label{rf7a}
\tilde{p} _{lm} =K_{lm} ^{l'm'} p_{lm}
\end{equation} 
with         
\[ 
p_{lm} =Y_{lm} (\zeta ,\bar{\zeta } )\frac{\partial }{\partial u}.
\]

The fourth order edth operator appearing in equation (\ref{rf6}) comes 
from the transformation properties of the quantities entering 
into the definition of 
 $\Psi $
; in particular, the term 
 $\eth^2 \bar\sigma$
 transforms 
under the law:
\begin{equation} \label {rf8}
\tilde{\sigma}^2 \tilde{\bar{\sigma } } =\frac{1}{K^{3} } \left( \eth^2  \bar{\sigma }- \eth^2 \bar \eth^2 \gamma \right);
\end{equation}
and similarly; after a straight forward but long calculation, 
one can prove that 
 $\Psi $
 transforms as
\begin{equation} \label{rf9}
\tilde{\Psi } =\frac{1}{K^{3} } \left( \Psi -\eth^{2}\bar \eth^{2} \gamma \right),
\end{equation}
where the transformation of Bondi systems under BMS transformations, 
appearing in the appendix of ref. \cite{Moreschi86}, have been used in deriving 
the last two relations. The nice section equation can be understood 
as a condition for 
 $\gamma (\zeta ,\bar{\zeta } )$
 and K.

It is important to note that there exists different choices for $\Psi $
 in the literature; because by adding any function with 
 $l\geq 2$
 (for 
example 
 $\eth^{2}\bar{\sigma } $
) one obtains the same expression for the Bondi momentum. 
But the supermomentum (\ref{rf2}) is the only one that have the simple 
transformation (\ref{rf9}). We mention also (but we do not make use of 
this) that the time derivative of (\ref{rf2}) have also a very simple 
form:
\[
\dot{\Psi } =\dot{\sigma } \dot{\bar{\sigma } }
;
\]
the other definitions of supermomentum have a flux that is not 
positive definite.

Although in equation (\ref{rf6}) one has an expression that involves 
the original Bondi coordinate system 
 $(u,\zeta ,\bar{\zeta } )$
, the difficulty with it 
is the appearance of the infinite transformation matrix 
 $K_{lm} ^{l'm'} $
; therefore 
it is convenient to seek a simpler expression. It was indicated 
in ref. \cite{Moreschi88} that equation (\ref{rf5}) can be expressed by
\begin{equation} \label{rf10}
\eth^2 \bar \eth^2 \gamma =\Psi
(u=\gamma ,\zeta ,\bar{\zeta } )+K(\gamma ;\zeta ,\bar{\zeta } )^{3}M(\gamma )
;
\end{equation}
where the function on the sphere $K$ is the conformal factor of 
the BMS transformation that aligns the 
 $\tilde{p} _{00} $
 with the Bondi momentum 
at 
 $u=\gamma (\zeta ,\bar{\zeta } )$
; which justifies the notation indicating the 
 $\gamma $
 dependence 
on it; and finally the total mass M is also calculated at 
 $\tilde{S} $
. Equation 
(\ref{rf10}) is our main equation that is studied throughout the rest 
of this article. The validity of equation (\ref{rf10}) can be deduced 
either from the discussion in ref. \cite{Moreschi88}, using the properties 
of the matrix 
 $K_{lm} ^{l'm'} $
; or it can be deduced from equations (\ref{rf5}) and (\ref{rf9}); 
since the nice section equation (\ref{rf5}) can be understood as requiring that $\tilde{\Psi } $
 be a constant when evaluated at the section $u=\gamma (\zeta ,\bar{\zeta } )$
; and this 
constant is precisely the negative of the Bondi mass M evaluated 
on this section.

Clearly $K(\gamma ;\zeta ,\bar{\zeta } )$ and 
 $M(\gamma )$ are determined by the value of the total momentum at $\tilde{S} $
, which can explicitly be given by:
\begin{equation} \label{rf11}
P^{\underline{a}} (\tilde{S} )=-\frac{1}{4\pi }
\int l^{\underline {a} } (\zeta ,\bar{\zeta } )\;\Psi (u=\gamma,\zeta ,\bar{\zeta } )\;dS^{2}
,
\end{equation}
where 
 $\underline{a} =0,1,2,3$
, and the vector 
 $l^{\underline{a} } $
 is given in terms of the first spherical 
harmonics by
\begin{equation} \label{rf12}
l^{\underline{a} } =\sqrt{4\pi } \left( Y_{00} ,-\frac{1}{\sqrt{6} }
(Y_{11} -Y_{1-1} ),\frac{i}{\sqrt{6} } (Y_{11} +Y_{1-1}
),\frac{1}{\sqrt{3} } Y_{10} \right)
.
\end{equation}
Note that one can also express the total momentum at 
 $\tilde{S} $
 by
\begin{equation} \label{rf13}
\begin{split}
P^{\underline {a}}( \tilde S)&= -\frac{1}{4\pi} \int l^{\underline {a}}(\zeta, \bar \zeta) \left (\int^\gamma_0 \dot \Psi(u', \zeta, \bar \zeta) du'+\Psi(u=0, \zeta, \bar \zeta)    \right) dS^2 \\
&=-\frac{1}{4\pi}\int l^{\underline {a}}(\zeta, \bar \zeta)\int^\gamma_0 \dot \Psi(u', \zeta, \bar \zeta)\, du'dS^2 +P^{\underline {a}}(S)\\
\end{split}
\end{equation}
where a dot means $u$ derivative; this shows a more explicit dependence 
of $P^{\underline{a} } (\tilde{S} )$ on 
 $\gamma (\zeta ,\bar{\zeta } )$
 and the momentum flux. Then, the conformal factor K can 
be given by
\begin{equation} \label{rf14}
K(\gamma ;\zeta ,\bar{\zeta } )=\frac{M(\gamma )}{P^{\underline{a} }
(\tilde{S} )l^{\underline{b} } (\zeta ,\bar{\zeta } )\eta _{\underline{a}\,
\underline{b} } },
\end{equation}

where the Lorentzian metric 
 $\eta _{\underline{a} \, \underline{b} } $
 was defined in \cite{Moreschi86} and 
\begin{equation} \label{rf15}
M(\gamma )=\sqrt{P^{\underline{a} } (\tilde{S} )P^{\underline{b} }
(\tilde{S} )\eta _{\underline{a} \,\underline{b} } }
.
\end{equation}

Then, it is observed that eq. (\ref{rf10}) is an elliptic, non linear, 
integro-differential equation for 
 $\gamma (\zeta ,\bar{\zeta } )$
. Let us mention same properties 
of this equation. Calling $D$ the differential operator appearing 
in the left hand side of equation (\ref{rf10}), we have that:
\begin{equation} \label{rf16}
D\equiv \eth^{2} \bar \eth^{2} =\frac{1}{4} \Delta ^{2} -\frac{1}{2} \Delta
\end{equation}

where 
 $\Delta $
 is the Laplacian of the unit sphere. Note that when the 
operator $D$ is applied to the spherical harmonics it gives\cite{Moreschi86}
\begin{equation} \label{rf17}
D Y_{lm} =\frac{\left( l-1\right) l\left( l+1\right) \left( l+2\right)
}{4} Y_{lm}
.
\end{equation}

This means that the left-hand side of (\ref{rf10}) has an expansion in 
spherical harmonics with 
 $l\geq 2$
; and in order to be a consistent equation 
the right-hand side must also have this property for an arbitrary $\Psi $ (see Appendix 1) for a proof of this property).

In the stationary case, 
 $\Psi =\Psi \left( \zeta ,\bar{\zeta } \right) $
, the equation becomes linear and has 
always solutions\cite{Moreschi88}. Let 
 $\gamma _{0} $
 be the a solution of the stationary 
case, then if 
 $x\left( \zeta ,\bar{\zeta } \right) $
 is such that: 
 $Dx=0$
 we have that 
 $\gamma _{0} +x$
 is also a solution. 
Then, in the stationary case there exists a 4-parameter family 
of solution (the real solution of 
 $Dx=0$
 are linear combinations of 
the spherical harmonics with 
 $l\leq 1$
, i.e.: they are determined by four 
real numbers). To illustrate this situation let us consider for 
example the Schwarzschild spacetime. In the usual coordinates 
and null tetrad the equation simplifies considerably since $\Psi =-M$ and $K=1$; where M is the Schwarzschild mass. Therefore, in this case, 
the right hand side of eq. (\ref{rf10}) vanishes, and so 
 $\gamma _{0} =0$
 is a nice section. 
This means that the usual coordinate system is a ``global rest 
frame'' system; and one still have the freedom of the translations 
\emph{x}'s. If instead we had begun with an arbitrary system in the 
Schwarzschild spacetime, then we would have 
 $\Psi \neq -M$
 and we would need 
a non trivial 
 $\gamma _{0} $
 to solve the nice section equation (\ref{rf10}). This $\gamma _{0} $ is the supertranslation we have to make in order to come back 
to ``global rest frame''. In order to gain the notion of center 
of mass from this family of solutions, one impose on the ``first'' 
section at 
 $u=-\infty $
 the condition that the angular momentum coincide 
with the intrinsic angular momentum; this eliminates a 3-dimensional 
freedom in the choice of the translations x's. Then, if from 
the first nice section we only allow for infinitesimal translations 
that are parallel to the total momentum, we generate from 
 $u=-\infty $
 the 
``center of mass system'' with the solutions of the nice section 
equation.

What happens in general, with a radiative asymptotically flat 
space-time?. Let put first a conjecture:

\emph{Conjecture:} Under mild hypothesis for 
 $\Psi $
, there exists  solutions 
 $\gamma $
 of (\ref{rf10}) 
of the form: 
 $\gamma =x+y$
, for each choice of translation x (
 $Dx=0$
), and with 
y a super-translation (i.e.: with spherical harmonics 
 $l\geq 2$
 expansion). 
The solution is 
 $C^{2} $
 in the sphere and if 
 $x_{1} $
 and 
 $x_{2} $
 are time constant 
translation, with 
 $x_{1} $
 in the future of 
 $x_{2} $
; that is:
\[
x_{1} \geq x_{2}
\]
then:
\[
\gamma _{1} \geq \gamma _{2}
(*)
\]
where $\gamma _{1} $ is the solution corresponding to $x_{1} $ and
$\gamma _{2} $ the corresponding to $x_{2} $.

If this conjecture were true, and by using the selection of ``initial'' 
nice section as it was explained for the Schwarzschild case, 
we would have that for any asymptotically flat space time there 
exist one globally defined ``rest frame'' at null infinity, that 
have the appropriate properties to be interpreted as the ``center 
of mass'' system of the space-time. We note that the last property 
(*), requires that the sections defining the center of mass system 
must be a one-parameter family, without self crossing behavior. 
We also remark that the differentiability property means that 
they are smooth sections of scri.

Of course this conjecture is very difficult to prove, because the
equation (\ref{rf10}) is a complicate non linear equation, but we can
give here an argument\cite{Moreschi88} (not a proof) that can help to
understand why one might think the conjecture is true.

The idea is to construct a perturbation series beginning with 
an arbitrary translation $x$. Let,
\[
\gamma _{0} =x \quad \gamma _{n} =x+\sum\limits_{k=1}^{n}y_{k}
\]
where $y_{n} $ is the solution of the linear equation
\[
\eth^2 \bar \eth^2 y_n =\Psi
(u=\gamma _{n-1} ,\zeta ,\bar{\zeta } )+K(\gamma _{n-1} ;\zeta ,\bar{\zeta
} )^{3} M(\gamma _{n-1} )
.
\]
If the series converges we have for each translation $x$ a solution
$\gamma $.

In the following part of this paper we prove the following results 
that indicate to the validity of the conjecture:

(\ref{rf1}) ``Local'' version of the conjecture: Given a space-time which 
satisfies, using geometric units, 
 $\left| \frac{\partial \Psi }{\partial u} \right| \leq \sqrt{\frac{27}{4}
} $
(**) and assuming there exists 
a solution 
 $\gamma _{0} $
, then there exists locally for each infinitesimal 
translation a solution 
 $\gamma $
, which is 
 $C^{2} $
 and satisfy (*). 

This is done in Sections III, IV, and V. Let us note that the 
condition (**) means that the spacetime does not admit arbitrary 
amount of radiation; however, for realistic astrophysical systems 
it turns out that (**) is always true (see \cite{Moreschi88}). 

(\ref{rf2}) If 
 $\Psi $
 is near to the stationary case, then the there exist a 
global 4-parameter family of solution.

This result is proved in Section VI.

\section{Local existence of solutions} \label{rfLocal}

The result of the local existence of solutions will be constructed 
around the implicit function theorem; therefore appropriate Banach 
spaces must be introduced. We begin by introducing some notation.

For any pair of regular functions on the sphere a and b let the 
inner product 
 $\langle a| b\rangle $
 be defined by:
\begin{equation} \label{rf18}
\langle a| b \rangle =\int \bar{a} b\;dS^{2},
\end{equation}
and associated to this let the norm of a function a be given 
by
\begin{equation} \label{rf19}
||a||^2 =\langle a| a \rangle ;
\end{equation}
that is; 
 $||\,||$ is the standard $L^2$ norm on the unit sphere 
 $S^{2}$.

We will next introduce three sets of real regular functions on 
the sphere. Let us start with the set:
\begin{equation} \label{rf20}
X=\left\{ x(\zeta ,\bar{\zeta } )\in L^{2} (S^{2} )/Dx=0
\right\};
\end{equation}
note that the real functions 
 $x(\zeta ,\bar{\zeta } )$
 are determined by four real numbers.

Let $Z$ be the set of real functions on the sphere with finite 
norm and residing in the orthogonal complement of $X$; equivalently
\begin{equation} \label{rf21}
Z=X^{\perp} =\{ z(\zeta ,\bar{\zeta } )\in L^2 (S^{2}
) / \langle x | z \rangle =0 \,   \forall    x\in X\}.
\end{equation}
Note that if $\gamma $ is in $L^{2} $
, then it can be decomposed as a sum of two 
terms; one in $X$ and the other in $Z$; that is 
 $\gamma =x+z$, $x\in X$, $z\in Z$. Furthermore, it 
is convenient to define the projector operator $T$, such that $T\left( \gamma \right) =T\left( x+z\right) =x$; 
that is T projects to the subspace $X$ and therefore satisfies $T^{2} =T$.

It is clear from the definitions that $X$ and $Z$ are Hilbert spaces.

Let us now define the following inner product. For any four times 
differentiable functions $y_{1} $ and $y_{2} $
, in $Z$, let $\langle | \rangle _{D} $
 be defined by
\begin{equation} \label{rf22}
\langle y_{1} | y_{2} \rangle _{D} \equiv \int
D\bar{y} _{1}Dy_{2} \;dS^{2}
;
\end{equation}
the fact that this is really an inner product in $Z$ follows from 
the existence of an inverse of the operator D; as it is deduced 
from equation (\ref{rf17}). It is natural to define the norm
\begin{equation} \label{rf23}
|| y||_{D}^{2} \equiv \langle y |
y \rangle _{D}.
\end{equation}

As it was stated in ref. \cite{Moreschi88} it is not difficult to see that
\begin{equation} \label{rf24}
|| y|| \leq ||y||_{D}.
\end{equation}

Let us define now the space of real functions on the sphere $Y$ 
by
\begin{equation} \label{rf25}
Y=\{ y(\zeta ,\bar{\zeta } )\in Z\;/\;|| y||_{D}
<\infty   \text{ completed with respect to this norm}\};
\end{equation}

If one expresses the functions $x$ and $y$ in terms of spherical  harmonics one has
\begin{equation} \label{rf26}
x=\sum^1_ {\substack{ l=0 \\ |m|\leq l}}
x^{lm} Y_{lm} (\zeta ,\bar\zeta )
\end{equation}
and
\begin{equation} \label{rf27}
y=\sum^\infty_{\substack{l=2 \\ |m| \leq l}} y^{lm} Y_{lm} (\zeta ,\bar \zeta ),
\end{equation}
for some constant coefficients $x^{lm} $ and $y^{lm} $.

Note that the space $Y$ is also a Hilbert space.

Given regular radiation data at future null infinity, let the real function $f(x,y)$ on the sphere be defined in terms of functions $x\in X$ and $y\in Y$ by the expression:
\begin{equation} \label{rf28}
f(x,y)=D{}{}y-\left( \Psi (u=x+y,\zeta ,\bar{\zeta } )+K(x+y;\zeta
,\bar{\zeta } )^{3} \;M(x+y)\right)
;
\end{equation}
where the radiation data enters into the determination of $\Psi $
, $K$ and $M$ at $u=\gamma =x+y$
; and it is easily seen that the condition $f(x,y)=0$ coincides with the nice section equation. It is also a property of any $f(x,y)$
, so constructed, that it lives in the orthogonal complement 
space of $X$; in other words, for any $f$ and any $x\prime $
 in $X$ one has 
 $\langle f|x'\rangle =0$; see Appendix  1 for a proof of this assertion. This means that 
$f$ is expressible in terms of spherical harmonics $Y_{lm}$ with $l\geq 2$; that 
is,
\begin{equation} \label{rf29}
f=\sum_{\substack{ l=2 \\| m| \leq l}} ^{\infty }f^{lm} \;Y_{lm} (\zeta ,\bar \zeta ).
\end{equation}
It is worth mentioning that the decomposition of the supertranslation $\gamma $ as the sum 
 $\gamma =x+y$
, is a natural one due to the fact that $Dx=0$
 for any 
$x$ in $X$.

Since 
 $\Psi $
, $K$ and $M$ are completely determined by the free data at 
future null infinity, one can see that for smooth data, any $f$ 
will have finite norm and so any $f$, so defined, belongs to $Z$.

Let us now consider the difference
\begin{multline*}
f(x,y+\delta y)-f(x,y)\\
=D \delta y-
\int\limits_{x+y}^{x+y+\delta y}\dot{\Psi } (u',\zeta ,\bar{\zeta } )\;du' 
+K(x+y+\delta y;\zeta ,\bar{\zeta } )^{3} \;M(x+y+\delta y)\\
-K(x+y;\zeta
,\bar{\zeta } )^{3} \;M(x+y)
\end{multline*}

From the definition (\ref{rf28}) one can prove that
\begin{equation} \label{rf30}
f(x,y+\delta y)-f(x,y)=f_{y} (x,y)\delta y+O(\delta y^{2} )
\end{equation}
with
\begin{multline} \label{rf31}
f_y (x,y)\delta y=D \delta y-\\
\left[ \dot{\Psi } (u=x+y,\zeta
,\bar{\zeta } )\delta y+K^{3} \left( \frac{4P_{\underline{a} } }{M}
-3K l_{\underline{a} } (\zeta ,\bar{\zeta } )\right) \delta
P^{\underline{a} } \right],
\end{multline}
where $P_{\underline{a}} $, $M$, and $K$ are evaluated at 
$u=x+y$, and $\delta P^{\underline{a} } $ is given by
\begin{equation} \label{rf32}
\delta P^{\underline{a} } =-\frac{1}{4\pi }
\int\limits_{}l^{\underline{a} } (\zeta ,\bar{\zeta } )\dot{\Psi }
(u=x+y,\zeta ,\bar{\zeta } )\;\delta y\;dS^{2};
\end{equation}
therefore by definition, $f_{y} (x,y)$
 is the Fr\'{e}chet derivative\cite{Berger77} of the map $f$ in terms of the argument $y$.

Let us observe that for smooth data $f_{y} (x,y)$ is continuous in $x$, and
\begin{equation} \label{rf33}
\delta z=f_{y} (x,y)\;\delta y
\end{equation}
is a linear continuous map from $Y$ onto $Z$. The \emph{onto} property is 
deduced from the fact that the first term in eq. (\ref{rf31}) is an operator 
providing an onto map.

Let us assume that
\begin{equation} \label{rf34}
f(x_{0} ,y_{0} )=0;
\end{equation}
that is, the pair $(x_{0} ,y_{0} )$ defines a nice section through the relation $\gamma _{0} =x_{0} +y_{0} $
, then
\begin{equation} \label{rf35}
P^{\underline{a} } (x_{0} ,y_{0} )=M(x_{0} ,y_{0} )\delta
_{0}^{\underline{a} },
\end{equation}
which implies that
\begin{equation} \label{rf36}
K(x_{0} ,y_{0} )=1.
\end{equation}

So the last term in equation (\ref{rf31}) is given by
\begin{equation} \label{rf37}
K^{3} \left( \frac{4P_{\underline{a} } }{M} -3K\;l_{\underline{a} }
(\zeta ,\bar{\zeta } )\right) \;\delta P^{\underline{a} } =\left(
4\;\delta _{\underline{a} }^{0} -3\;l_{\underline{a} } (\zeta ,\bar{\zeta
} )\right) \;\delta P^{\underline{a} };
\end{equation}
which is clearly an expression involving only spherical harmonics $Y_{lm} $ with $l\leq 1$.

Therefore one has that(see also Appendix 1)
\begin{equation} \label{rf38}
f_{y} (x_{0} ,y_{0} )\delta y=D{}{}\delta y-\left[ F\;\delta
y-T(F\delta y)\right],
\end{equation}
with $F=F( \zeta ,\bar{\zeta }) \equiv \dot{\Psi } ( u=x_{o}
+y_{o} ,\zeta ,\bar{\zeta }) $. As it was done in ref. \cite{Moreschi88}, we will assume that the modulus 
of the flux is bounded by the quantity $\lambda $, i.e.:
\begin{equation} \label{rf39}
\left| \dot{\Psi } (u,\zeta ,\bar{\zeta } )\right| \leq \lambda;
\end{equation}
and therefore $\left| F\right| \leq \lambda $.

Then the following result holds:

\begin{lemma}
If $\left( \lambda ^{2} \frac{4}{27} \right) <1$ then the map $\delta z=f_{y} (x_{0} ,y_{0} )\delta y$
 from $Y$ onto $Z$ is invertible.
\end{lemma}
\emph{Proof}:We start with the expression
\begin{equation} \label{rf40}
D\delta y-\left[ F\delta y-T(F\delta y)\right] =\delta z.
\end{equation}
Note that from eq. (\ref{rf17}) one can see that there exists an operator $D^{-1} $, acting on elements of Z, such that on spherical harmonics, with  $l\geq 2$
, it gives
\begin{equation} \label{rf41}
D^{-1}Y_{lm} =\frac{4}{( l-1) l( l+1) (
l+2) } Y_{lm},
\end{equation}
then one deduces that  $z\in Z\Rightarrow D^{-1} z\in Z$. Let us now define the operator  $A:Y\rightarrow Y$
, such that  $\forall \delta z\in Z$,
\begin{equation} \label{rf42}
A\delta y\equiv D^{-1} ( \delta z+F\delta y-T( F\delta
y ).
\end{equation}
Then eq. (\ref{rf40}) is equivalent to the equation:
\begin{equation} \label{rf43}
A\delta y=\delta y.
\end{equation}
We will next show that if $\left( \lambda ^{2} \frac{4}{27} \right) <1$ then $A$ is a contracting map\cite{Berger77}, and 
therefore eq.(\ref{rf43}) has a unique solution.

It is shown in Appendix 2 that for any function 
 $\Gamma (\zeta ,\bar{\zeta } )\in Y$
 one has the 
following inequality:
\begin{equation} \label{rf44}
|| \Gamma|| ^2 \leq \frac{4}{27} ||\Gamma ||_D^{2}.
\end{equation}
This clearly improves the inequality (\ref{rf24}). It then follows that
\begin{equation} \label{rf45}
\begin{split} || A \delta y- A \delta y'|| &= || D^{-1} F(\delta y-\delta y')-D ^{-1}T(F(\delta y-\delta y'))||^2_D\\
&= ||F(\delta y-\delta y')-T(F(\delta y-\delta y'))||^2\\
&\leq ||F(\delta y-\delta y')||^2\leq \lambda^2||(\delta y-\delta y')||^2\\
&\leq \lambda^2\frac{4}{27}||(\delta y-\delta y')||^2_D,
\end{split}
\end{equation}
where we have used the fact that for any $\gamma \in L^{2} ( S^{2}) $ one has  $|| \gamma -T\gamma || \leq || \gamma || $. Consequently 
$A$ defines a contracting map, which implies that (\ref{rf43}) has a unique 
solution. Then $f_{y} (x_{0} ,y_{0} )$
 provides with a one-to-one map and onto from 
$Z$ to $Y$.

Therefore  $f_{y} (x_{0} ,y_{0} )$ is an homeomorphism of $Y$ onto $Z$. $\blacksquare$

This result along with the implicit function theorem\cite{Berger77} allows 
us to prove the following:

\begin{theorem}
Let  $\gamma _{0} =x_{0} +y_{0} $ be a nice section, i.e.: 
 $f(x_{0} ,y_{0} )=0$
, and the smooth data 
 $\dot{\Psi } $
 satisfying  $| \dot{\Psi } (u,\zeta ,\bar{\zeta } )| \leq \lambda $
, with 
 $\lambda <\sqrt{\frac{27}{4} } $
, then there exist a local 4-parameter family of nice 
sections near 
 $\gamma _{0} $.
\end{theorem}
\emph{Proof:} Lemma 3.1 allows us to use the implicit function theorem\cite{Berger77} for 
the Banach spaces $X$, $Y$ and $Z$, and the map 
 $z=f(x,y)$
; which tell us that 
there is a unique continuous mapping 
 $g:U\rightarrow Y$
 defined in a neighborhood 
$U$ of 
 $x_{0} $
 such that 
 $y_{0} =g(x_{0} )$
 and
\begin{equation} \label{rf46}
D g(x)=\Psi (u=x+g(x),\zeta ,\bar{\zeta } )+K\left( x+g(x);\zeta
,\bar{\zeta } \right) ^{3} M\left( x+g(x)\right).
\end{equation}

Since there is a 4-parameter family of solutions to the equation  $Dx=0$
; one deduces that given a nice section and data satisfying 
 $\lambda <\sqrt{\frac{27}{4} } $, 
there exists a 4-parameter family of them in a neighborhood of 
the original one.$\blacksquare$

This result generalizes the calculations of ref. \cite{Moreschi88} in several 
ways; notably by pushing the permissible upper value of 
 $\lambda $
 from $1$ to 
 $\sqrt{\frac{27}{4} } $
, and also by allowing a general translation $x$.

\section{Differentiable properties of the local solutions} \label{rfDifferentiable}

In order to study the differentiable properties of nice sections 
we need to introduce further notation. Let us express the Sobolev norm  $H_{4} $ by\cite{Aubin82}:
\begin{equation} \label{rf47}
\left\| y\right\| _{H_{4} } =\sum\limits_{i=0}^{4}\left\| \nabla ^{(i)}
y\right\|;
\end{equation}
where the multi-index $i$ is used as
\begin{equation} \label{rf48}
| \nabla ^{(i)} y| ^{2} =\nabla ^{a_{1} } \nabla ^{a_{2} }
\cdots \nabla ^{a_{i} } y\nabla _{a_{1} } \nabla _{a_{2} } \cdots \nabla _{a_{i}
} y.
\end{equation}

Since the operator $D$ is strongly elliptic\cite{Friedman69} of fourth degree on the compact manifold  $S^{2} $
, then the following inequalities hold\cite{Rendall88}\cite{Choquet81}:
\begin{equation} \label{rf49}
\left\| Dy\right\| \leq c_{1} \left\| y\right\| _{H^{4} },
\end{equation}

\begin{equation} \label{rf50}
\left\| y\right\| _{H^{4} } \leq c_{2} \left( \left\| Dy\right\|
+\left\| y\right\| \right),
\end{equation}
for some positive constants  $c_{1} $ and  $c_{2} $.

We will use these relations to prove the next result.

\begin{lemma}
There exist constants $C$ and 
 $C' $
, independent of $y$ such that  $\forall y\in Y$
 one 
has that:
\begin{equation} \label{rf51}
C|| y|| _{H^{4} } \leq || y|| _{D} \leq C'
|| y|| _{H^{4} }.
\end{equation}

That is the norms are equivalent.
\end{lemma}
\emph{Proof:} The second inequality is just relation (\ref{rf49}). The first inequality 
follows from (\ref{rf50}) and the fact that for  $y\in Y$
 one has:
\begin{multline}
|| y|| _{H^{4} } \leq c\left( || Dy|| +||
y|| \right) =c\left( || y||_{D} +|| y||
\right)\\ 
\leq c\left( || y|| _{D} +\sqrt{\frac{4}{27} }
|| y||_{D} \right) =c\left( 1+\sqrt{\frac{4}{27} } \right)
|| y|| _{D}.
\end{multline} $\blacksquare$

It then follows:
\begin{theorem}
The local 4-parameter family of nice sections is two times differentiable 
functions on the sphere.
\end{theorem}
\emph{Proof:} Using the last Lemma and the Sobolev imbedding theorem for compact 
manifolds(see for example \cite{Aubin82}) one deduces that the solutions 
$g(x)$, mentioned in the proof of the theorem of the last section, 
are in $C^{2} (S^{2} )$
. $\blacksquare$

\section{Non self-crossing property of the local solutions}
\label{rfNon self}
Let  $\gamma (\tau )$ be a one parameter family of nice sections for  $\tau \in [ 0,t]$
. Then since $\gamma (x,y)$ is linearly expressed in terms of $x$ and $y$ by  $\gamma (\tau )=x(\tau )+y(x(\tau ))$
, it is clear that 
the Fr\'{e}chet derivative of  $f(\gamma (x,y))$
, with respect to  $\gamma $, 
 $f_{\gamma } (\gamma (x,y))$
, is given by the analog 
expression (\ref{rf31}) when one substitutes $\delta y$ by 
 $\delta \gamma $
; that is:
\begin{multline} \label{rf52}
f_{\gamma } (\gamma (x,y))\delta \gamma =D\delta \gamma \\
-\left[
\dot{\Psi } (u=\gamma ,\zeta ,\bar{\zeta } )\;\delta \gamma +K^{3} \left(
\frac{4P_{\underline{a} } }{M} -3K\;l_{\underline{a} } (\zeta ,\bar{\zeta
} )\right) \delta P^{\underline{a} } (\delta \gamma )\right],
\end{multline}
where $P_{\underline{a} }$, $M$, and $K$ are evaluated at 
 $u=\gamma $
, and 
 $\delta P^{\underline{a} } (\delta \gamma )$
 is given by
\begin{equation} \label{rf53}
\delta P^{\underline{a} } (\delta \gamma )=-\frac{1}{4\pi }
\int\limits_{}l^{\underline{a} } (\zeta ,\bar{\zeta } ) \dot{\Psi }
(u=\gamma ,\zeta ,\bar{\zeta } ) \delta \gamma \;dS^{2} 
.
\end{equation}

Recalling that the nice section equation is equivalent to 
 $f(x,y)=0$
, it 
is deduced from the properties of the equation, that its Fr\'{e}chet derivative with respect to 
 $\gamma $
 must also vanish. But it is 
observed that the equation 
 $f_{\gamma } (\gamma )\;\delta \gamma =0$
 coincides with the linearization 
of the nice section equation; therefore, if one takes 
 $\delta \gamma =\frac{d\gamma }{d\tau } \;d\tau $
 and wants 
to study the properties of 
 $\frac{d\gamma }{d\tau } $
, one needs to analyze the linearized 
nice section equation.

So, let us consider now the solutions of the linear nice section 
equation of the form
\begin{equation} \label{rf54}
\delta \gamma =\delta x+\Gamma,
\end{equation}
with  $\delta x\in X$ and  $\Gamma \in Y$
. From the arguments of ref.\cite{Moreschi88}, and their extension 
to general translations, presented above, it is deduced that 
a solution can be constructed by the sequence 
 $\delta \gamma _{n} =\delta x+\Gamma _{n} $
 defined in the 
following form:
\begin{equation} \label{rf55}
\delta \gamma _{0} =\delta x
\end{equation}

\begin{equation} \label{rf56}
\delta \Gamma _{1} =A_{0}\delta \gamma _{0},
\end{equation}
and in general, for  $n>1$,
\begin{equation} \label{rf57}
\delta \Gamma _{n} =A_{0}(\delta x+\delta \Gamma _{n-1} )
\end{equation}
or
\begin{equation} \label{rf58}
\delta \gamma _{n} =\delta x+A_{0}\delta \gamma _{n-1},
\end{equation}
where  $A_{0} $
 is the trivial extension of the operator defined in (\ref{rf42}) 
to act on  $X\oplus Y$
, with $\delta z=0$. Let us note that if 
 $\delta x=0$, then  $\delta \gamma _{1} =0$
 which implies that  $\delta \gamma _{n} =0\;\forall n$
, or equivalently  $\Gamma =0$.

It is interesting to observe that this solution for the homogenous 
linear equation can also be expressed by
\begin{equation} \label{rf59}
\delta \gamma =\sum_{i=0}^{\infty }A_{0} ^{i} \delta x;
\end{equation}
in particular, one can see that
\begin{equation} \label{rf60}
\Gamma _{n} =\sum_{i=1}^{n} A_{0} ^{i} \delta x .
\end{equation}

The proof that (\ref{rf59}) gives the solution of the homogeneous linear 
equation can be given in the following way. First let us check 
that if the series (\ref{rf59}) converges then it is a solution, since
\begin{equation} \label{rf61}
A_{0} \delta \gamma =A_{0} \left( \sum_{i=0}^{\infty }A_{0}
^{i} \delta x \right) =\sum_{i=1}^{\infty }A_{0} ^{i} \delta x
=\Gamma ;
\end{equation}
which is the representation of the homogeneous linear equation 
in terms of the operator  $A_{0} $. Now let us prove that the series (\ref{rf59}) 
converges. Note that
\begin{equation} \label{rf62}
\left\| \delta \gamma _{n} -\delta \gamma _{n-1} \right\| _{D} =\left\|
\Gamma _{n} -\Gamma _{n-1} \right\| _{D} =\left\| A_{0} ^{n} \;\delta
x\right\| _{D};
\end{equation}
then, since for  $n\geq 1$,  $A_{0} ^{n}\delta x\in Y$
, one has from (\ref{rf45}) that
\begin{equation} \label{rf63}
\left\| A_{0} y\right\| _{D} \leq \lambda \;\sqrt{\frac{4}{27} }
\;\left\| y\right\| _{D}
;
\end{equation}
therefore
\begin{equation} \label{rf64}
\left\| \delta \gamma _{n} -\delta \gamma _{n-1} \right\| _{D} \leq
\left( \lambda \;\sqrt{\frac{4}{27} } \right) ^{n-1} \left\| A_{0}
\delta x\right\| _{D}
;
\end{equation}
which implies that the sequence converges if 
 $\left( \lambda \sqrt{\frac{4}{27} } \right) <1$.

Using the notation of the last section, and denoting 
 $\dot{\Psi } _{0} =\dot{\Psi } (u=x_{0} +g(x_{0} ),\zeta ,\bar{\zeta } )$
, one can 
prove the following result:
\begin{theorem}
Let  $\gamma _{0} =x_{0} +g(x_{0} )$ be a nice section and let  $\delta x_{0} $ be a positive constant 
such that 
 $x=x_{0} +\delta x_{0} \in U$
. There exist  $\lambda _{*} \in (0,\sqrt{\frac{27}{4} } )$
 such that for data satisfying 
 $| \dot{\Psi } _{0} | <\lambda _{*} $
 the solution  $\delta \gamma (\delta x_{0} )$
 of the linear nice section equation is also positive.
\end{theorem}

This implies, as is mentioned below, that the solutions of the 
nice section equation generated by an aligned time translation 
do not cross among them.

\emph{Proof:} Using the fact that for any  $\gamma \in L^{2} \left( S^{2} \right) $,  $\left\| \gamma -T\gamma \right\| \leq \left\| \gamma \right\| $
, one deduces that if 
 $\delta \gamma =\delta x+\Gamma $
 is a solution 
of the linear nice section equation, then

\begin{align*}
||\Gamma || _{D}^{2}  & \leq  || \dot{\Psi } _{0}
(\delta x+\Gamma )|| ^{2} \leq \lambda ^{2} \ \left\| \delta
x+\Gamma \right\| ^{2} \leq \lambda ^{2} \ \left\| \ \Gamma \right\| ^{2}
+\lambda ^{2} \ \left\| \ \delta x\right\| ^{2}  \\
& \leq  \lambda ^{2} \left( \frac{4}{27} \left\| \Gamma \right\|
_{D}^{2} +\ \left\| \ \delta x\right\| ^{2} \right); 
\end{align*}
therefore
\begin{equation} \label{rf66}
||\Gamma || _{D}^{2} \leq \frac{\lambda ^{2}
}{1-\frac{4}{27} \lambda ^{2} } || \delta x|| ^{2}.
\end{equation}

From Appendix 2, one has
\begin{equation} \label{rf67}
||\Gamma (\zeta ,\bar{\zeta } )|| ^{2} \leq \frac{1}{4\pi }
\;\frac{4}{27} ||\Gamma || _{D}^{2};
\end{equation}
therefore, it is deduced that
\begin{equation} \label{rf68}
\sup \left| \Gamma \right| ^{2} <\frac{4}{27} \frac{\lambda ^{2}
}{1-\frac{4}{27} \lambda ^{2} } \left( \delta x_{0} \right) ^{2}.
\end{equation}

Then, the inequality
\begin{equation} \label{rf69}
\sup \left| \Gamma \right| ^{2} <\left( \delta x_{0} \/\right) ^{2} ;
\end{equation} 
will be satisfied if
\begin{equation} \label{rf70}
\lambda <\lambda _{*} =\frac{1}{\sqrt{2} } \sqrt{\frac{27}{4} } \cong
1.8371
\end{equation} 
$\blacksquare$

The constant 
 $\lambda _{*} $
 appearing in the result depends on the technique 
one uses for the proof. In order to illustrate this, in the Appendix 
3 we present three other alternative proofs, using different 
techniques, which impose three different bounds on the data; 
which do not improve the last one.

It is not yet clear to us whether the above value of 
 $\lambda _{*} $
 is the 
best possible one.

Theorem 5.1 proves that 
 $\delta \gamma >0$
 for an aligned future time translation 

 $d\tau $
, which implies that 
 $\frac{d\gamma }{d\tau } >0$
, and therefore that the nice section family 
does not show self crossing since 
 $\gamma (\tau +d\tau )$
 is to the future of 
 $\gamma (\tau )$
.
\section{Global existence of nice sections near the stationary case} \label{rfGlobal}

The results appearing in the previous sections deal only with local
properties of nice sections; in this section the issue of global existence
will be tackled. In Ref. 2 it was proved that there exists a 4-parameter
family of nice sections in a stationary spacetime. This suggests that when
radiation is ``small'' one could probably find an analogous result. In what
follows we will apply once more the implicit function theorem to prove this
assertion.

Let $C^{0}({\cal I}^{+})$ represent the vector space of all bounded
continuous functions on scri plus, with the norm $\left\| F\right\| _{C^{0}(%
{\cal I}^{+})}=\sup \left| F(u,\zeta ,\overline{\zeta })\right| $. The
vector space $C^{0}({\cal I}^{+})$ defines a Banach space.

Let $X$, $Y$ and $Z$ represent the spaces defined in Sec. III, and let us
denote with $\Xi $ the space whose elements are of the form $\xi =(x,F)$ ,
with $x\in X$, $F\in C^{0}({\cal I}^{+})$ and with norm $\left\| \xi
\right\| =\left\| x\right\| _{X}+\left\| F\right\| _{C^{0}({\cal I}^{+})}$.
The space $\Xi $ is a Banach space. Consider now the map from $(\Xi ,Y)$
into $Z$ defined by
\begin{multline} \label{rf71}
\Phi (\xi ,y)=Dy-\\
\left( \int\limits_{u_{0}}^{x+y}F(u,\zeta ,\bar{\zeta}%
)\;du+\Psi (u_{0},\zeta ,\overline{\zeta })+K(x+y,F;\zeta ,\bar{\zeta}%
)^{3}\;M(x+y,F)\right) 
\end{multline}
which coincide with the map $f(x,y)$ defined in Sec. III, but where now it
is taken into account explicitly the dependence on the flux $F$. The
Fr\'{e}chet derivative of this map with respect to the variable $y$ is
\begin{multline} \label{rf72}
\Phi _{y}(\xi ,y)\delta y=D \delta y\\
-\left[ F(u=\gamma =x+y,\zeta ,\bar{
\zeta})\;\delta y+K^{3}\left( \frac{4P_{\underline{a}}}{M}-3K\;l_{\underline{%
a}}(\zeta ,\bar{\zeta})\right) \delta P^{\underline{a}}\right] 
\end{multline}
where $P_{\underline{a}}$ , $M$, and $K$ are evaluated at $u=\gamma =x+y$ ;
in particular one has:
\begin{equation} \label{rf73}
P^{\underline{a}}(\gamma )=-\frac{1}{4\pi }\int\limits_{{}}l^{\underline{a}%
}(\zeta ,\bar{\zeta})\int\limits_{u_{0}}^{\gamma }F(u,\zeta ,\bar{\zeta}%
)du\;dS^{2}+P^{\underline{a}}(u=u_{0})\quad (\ref{rf73}) 
\end{equation}
and $\delta P^{\underline{a}}$ is given by
\begin{equation} \label{rf74}
\delta P^{\underline{a}}=-\frac{1}{4\pi }\int\limits_{{}}l^{\underline{a}%
}(\zeta ,\bar{\zeta})\;F(u=\gamma =x+y,\zeta ,\bar{\zeta})\;\delta
y\;dS^{2}.
\end{equation}
Let us denote with $\xi _{0}$ the non-radiation case, that is, $\xi
_{0}=(x_{0},F=0)$; we have shown in Ref. 2 that there exists a relation $%
y_{0}(x_{0})$ such that $\Phi (\xi _{0},y_{0})=0$, for any translation $%
x_{0} $; in other words, in the absence of radiation $(F=0)$, the map $%
\gamma _{0}(x_{0})=x_{0}+y_{0}(x_{0})$ defines a 4-parameter family of nice
sections globally on scri. It is observed that $\delta P^{\underline{a}%
}(F=0)=0$ ; therefore
\begin{equation} \label{rf75}
\Phi _{y}(\xi (x,F=0),y)\delta y=D\hspace{0.1cm}\delta y, 
\end{equation}
and it is clear that the map $\delta z=\Phi _{y}(\xi (x,F=0),y)\delta y$
from $Y$ onto $Z$ is invertible; that is, it is an homeomorphism of $Y$ onto 
$Z$.

The implicit function theorem implies that there exists a unique continuous
mapping $G$ defined in a neighborhood $U_{0}\subset C^{0}({\cal I}^{+})$ of $%
F=0$, $G:U_{0}\rightarrow Y$, parametrized by $x_{0}$, such that $G(\xi
(x_{0},F=0))=y_{0}$ and $\Phi (\xi (x_{0},F),y)=0$ for $F\in U_{0}$.

Let us call ``small radiation data'' the one that satisfies
$\dot\sigma\dot{\bar\sigma}=F\in U_{0}$ in the last result. Then, the
continuity properties contained in the implicit function theorem and
the fact that $x$ lives in a 4-dimensional vector space, imply the
following result:
\begin{theorem}
For ``small radiation data'' there exists a global 4-parameter family of
nice sections at scri.
\end{theorem}

\section{Final comments}

Summarizing, in this article several results concerning local and
global properties of the center of mass of isolated systems in general
relativity have been presented; in terms of the so called nice
sections. In Section \ref{rfGlobal} the global existence of nice
sections has been established for small radiation data. In Section
\ref{rfLocal} there appears a formal proof of existence of nice
sections with less restrictive assumptions on the strength of the
radiation data that had appear in the literature before[2].

Although the global result of Section \ref{rfGlobal} is very important from 
a formal perspective of the whole picture based on nice sections; 
the local result of Section \ref{rfLocal} might be more important from 
the operational point of view, since if one wants to make a concrete 
calculation based on this construction it is natural to start 
by taking an initial nice section and calculating then the dynamics 
of the system on a local family around the initial one. The relaxation 
of the strength of the radiation warrants a wider set of applicability 
of this formalism.

Differentiable properties of nice sections have been established in
Section \ref{rfDifferentiable}. And in Section \ref{rfNon self} the
desired non-self-crossing property of nice sections was proved.

It is important to emphasize that this last property is what 
one expects from making the analogy with the case of a set of 
particles in linearized gravity; since in this case any sensible 
notion of center of mass, defined in the interior of the spacetime, 
would follow a timelike curve whose future null cones would define 
the corresponding one parameter family of center of mass at future 
null infinity. This later family clearly could not show self-crossing 
since this would imply a faster than light velocity for the trajectory 
of the center of mass in the interior of the spacetime.

We conclude then that the present approach for the notion of 
center of mass of isolated gravitating systems, based on the 
definition of nice sections, not only has the necessary mathematical 
properties, but it also has the appropriate physical properties 
that make possible its interpretation.

\section*{Acknowledgments}

Most of this work was developed while one of the authors (O.M.)  was
visiting the Albert-Einstein-Institut, Max-Planck-Institut f\"{u}r
Gravitationsphysik, Potsdam, Germany. This author is very grateful to
all the people at the Institute and in particular Dr. Helmut
Friedrich, for their kind hospitality.

It is a pleasure to thank Drs. H. Friedrich, O. Reula, B. Schmidt and
E. T. Newman for illuminating discussions on our work. We also would
like to thank A. Rendall for pointing out an error in the original
paper and for discussions.

This research was supported by the ``Consejo de Investigaciones
Cient\'{\i}ficas y Tecnol\'{o}gicas de la Provincia de C\'{o}rdoba''
(CONICOR), the ``Consejo Nacional de Investigaciones Cient\'{\i}ficas
y T\'{e}cnicas'' (CONICET) and the ``Deutsche Forschungsgemeinschaft''
(DFG).

\section*{Appendix 1} \label{rfAppendix 1}

For any regular real function on the sphere f let us define
\[
f^{\underline{a} } =\frac{1}{4\pi } \int l^{\underline{a} } (\zeta
,\bar{\zeta } )\;f _{}^{} (\zeta ,\bar{\zeta } )\;dS^{2} , 
\]
and let us also define
\[
\hat{f} \equiv f^{\underline{a} } \;l_{\underline{a} } =f^{0} -\hat{f}
^{(\ref{rf3})},
\]
with  $\hat{f} ^{(\ref{rf3})} \equiv ( f^{1} l^{1} +f^{2} l^{2} +f^{3} l^{3}
) $
. Then it is easy to prove by direct calculation
\begin{lemma}
\[
T(f)=f^{0} -3\;\hat{f} ^{(\ref{rf3})}.
\]
\end{lemma}
It is also useful the following result.

\begin{lemma}
For any timelike vector  $U^{a} $ one has 
\[
\frac{1}{4\pi } \int \frac{l^{\underline{a} } }{\left( U^{\underline{b}
} l_{\underline{b} } \right) ^{3} } dS^{2}  =\frac{U^{\underline{a} }
}{\left( U^{\underline{b} } U_{\underline{b} } \right) ^{2} }
.
\]
\end{lemma}
\emph{Proof:}

First let us prove that the left hand side transforms as a 4-vector 
under Lorentz transformations. This can easily be seen from the 
fact that under Lorentz transformations:
\[
l^{\underline{a} } \rightarrow K l^{\underline{a} }, \quad dS^{2}
\rightarrow K^{2} dS^{2} .
\]
Therefore
\[
\frac{1}{4\pi } \int \frac{l^{\underline{a} } }{\left( U^{\underline{b}
} l_{\underline{b} } \right) ^{3} } dS^{2}  =\frac{V^{\underline{a} }
}{\left( U^{\underline{b} } U_{\underline{b} } \right) ^{2} }
;
\]
for some 4-vector $V^{\underline{a} } $

Since  $U^{a} $
 is a timelike vector, one can choose a frame for which
\[
U^{\underline{a} } =\left( U^{0} ,0,0,0\right);
\]
from which it is immediately deduced that
\[
V^{\underline{a} } =\left( U^{0} ,0,0,0\right) =U^{\underline{a} }
.
\]
$\blacksquare$

Then it follows:
\begin{corollary}
Let  $U^{a} $ be a timelike vector, then:
\[
T\left( \frac{1}{\left( U^{\underline{a} } l_{\underline{a} } \right)
^{3} } \right) =\frac{1}{\left( U^{\underline{a} } U_{\underline{a} }
\right) ^{2} } \left( u^{0} -3\hat{u} ^{(\ref{rf3})} \right);
\]
where:
\[
U^{\underline{a} } \;l_{\underline{a} } =u^{0} -\hat{u} ^{(\ref{rf3})}
.
\]
\end{corollary}
\emph{Proof:}

From Lemma A1.1, one has:
\begin{align*}
T\left( \frac{1}{(U^{\underline{a}} l_{\underline{a}})^3 }\right)&=\frac{1}{4\pi} \int \frac{1}{(U^{\underline{a}} l_{\underline{a}})^3} (l^0 l_0 - 3 l^i(\zeta', \bar \zeta'))l^i (\zeta, \bar \zeta)){dS'}^2\\
&=\frac{1}{(U^{\underline{a}}U_{\underline{a}})^2}(u^0-3 {\hat u}^{(\ref{rf3})}), 
\end{align*}
where i=1,2,3 . $\blacksquare$

Then it follows that
\[
T(\Psi )=-T(K^{3} M).
\]

\section*{Appendix 2} \label{rfAppendix 2}

Let us note that for any function $\Gamma (\zeta ,\bar{\zeta } )\in Y$ one has

\begin{align*}
  | \Gamma (\zeta ,\bar{\zeta } )| ^{2} & = \left|
    \sum_{l=2}^{\infty }\sum_{m=-l}^{l} \Gamma ^{lm} \ Y_{lm}(\zeta ,\bar{\zeta } ) \right| ^{2}  \\
  & = \left| \sum_{l=2}^{\infty }\sum_{m=-l}^{l} \Gamma
    ^{lm}  \frac{(l-1)l(l+1)(l+2)}{4} \frac{4}{(l-1)l(l+1)(l+2)} Y_{lm}(\zeta ,\bar{\zeta } )  \right| ^{2}  \\
  & \leq \sum\limits_{l=2}^{\infty }\sum\limits_{m=-l}^{l}\ \left|
    \Gamma ^{lm} \right| ^{2} \ \left[ \frac{(l-1)l(l+1)(l+2)}{4}
  \right] ^{2} \sum\limits_{l=2}^{\infty }\sum\limits_{m=-l}^{l}\ 
  \left[ \frac{4\ }{(l-1)l(l+1)(l+2)} \right] ^{2} \left| Y_{lm}
    (\zeta ,\bar{\zeta } )\right| ^{2}
\end{align*}

Now, using the addition theorem for the spherical harmonics, 
one can prove that
\[
\sum\limits_{m=-l}^{l}\;\left| Y_{lm} (\zeta ,\bar{\zeta } )\right| ^{2}
\; \leq \frac{2l+1}{4\;\pi }
.
\]

Also one can see\cite{Moreschi86} that
\[
\left\| \Gamma \right\| _{D}^{2} =\left[ \sum\limits_{l=2}^{\infty
}\sum\limits_{m=-l}^{l}\;\left| \Gamma ^{lm} \right| ^{2} \;\left[
\frac{(l-1)l(l+1)(l+2)}{4} \right] ^{2}   \right]
\]
from which it is deduced that
\[
\left| \Gamma \right| ^{2} \leq \frac{1}{4\pi } \left\| \Gamma \right\|
_{D}^{2} \;\left\{ \sum\limits_{l=2}^{\infty }\;\left[
\frac{4\;\;}{(l-1)l(l+1)(l+2)\;} \right] ^{2} (2l+1) \right\}
=\frac{1}{4\pi } \;\frac{4}{27} \;\left\| \Gamma \right\| _{D}^{2}.
\]

Since the modulus  $|| \Gamma || $ is bounded by its maximum value, namely
\[
\left\| \Gamma \right\| ^{2} \leq 4\;\pi \;\;\sup \left| \Gamma (\zeta
,\bar{\zeta } )\right| ^{2};
\]
it is deduced the inequality:
\begin{equation} \label{B1}
\left\| \Gamma \right\| ^{2} \leq \frac{4}{27} \left\| {}\Gamma \right\|
_{D}^{2}.
\end{equation}

\section*{Appendix 3} \label{rfAppendix 3}

Since the peculiar value of  $\lambda _{*} $
 depends on the technique of the 
proof of Theorem 5.1; we present here three other proofs of this 
result.

\emph{Proof 1:} Let us observe that
\[
\langle \Gamma | \eth^2 \bar \eth^2 \Gamma \rangle =\langle \bar
\eth^2 \Gamma | \bar \eth^2 \Gamma \rangle =\langle \eth^2 \Gamma |
\eth^2 \Gamma \rangle;
\]
the question arises then whether there is an extension of the 
Sobolev inequality that will relate 
 $\left| \Gamma (\zeta ,\bar{\zeta } )\right| $
 with the previous expression. 
We will now prove that there is such a relation.

Let the regular function  $\Gamma $
 be given by
\[
\Gamma =\sum_{l=2}^{\infty }\sum_{m=-l}^{l}\Gamma ^{lm}
Y_{lm} (\zeta ,\bar{\zeta } ),
\]
then one has

\begin{align*}
| \Gamma (\zeta ,\bar{\zeta } )| ^{2}  & =  \left|
\sum\limits_{l=2}^{\infty }\sum\limits_{m=-l}^{l}\ \Gamma ^{lm} \ Y_{lm}
(\zeta ,\bar{\zeta } )  \right| ^{2}  \\
& =  \left| \sum\limits_{l=2}^{\infty }\sum\limits_{m=-l}^{l}\ \Gamma
^{lm} \ \sqrt{\frac{(l-1)l(l+1)(l+2)}{4} }
\sqrt{\frac{4\ }{(l-1)l(l+1)(l+2)} } Y_{lm} (\zeta ,\bar{\zeta } ) 
\right| ^{2}  \\
& \leq   \sum_{l=2}^{\infty }\sum_{m=-l}^{l} \left|
\Gamma ^{lm} \right| ^{2}  \left[ \frac{(l-1)l(l+1)(l+2)}{4} \right] 
\sum_{l=2}^{\infty }\sum_{m=-l}^{l}\ \left[
\frac{4\ }{(l-1)l(l+1)(l+2)} \right] ^{} \left| Y_{lm} (\zeta ,\bar{\zeta
} )\right| ^{2}
\end{align*}

Now, using the inequality implied by the addition theorem for 
the spherical harmonics, and that\cite{Moreschi86}
\[
\langle \Gamma |\eth^2 \bar \eth^2\Gamma \rangle = \sum\limits_{l=2}^{\infty
}\sum\limits_{m=-l}^{l}\;\left| \Gamma ^{lm} \right| ^{2} \;\left[
\frac{(l-1)l(l+1)(l+2)}{4} \right]
\]
one deduces
\[
|\Gamma| \leq \frac{1}{4\pi} \langle \Gamma | \eth^2 \bar \eth^2
\Gamma \rangle \sum_{l=2}^\infty
\frac{4(2l+1)}{(l-1)l(l+1)(l+2)}=\frac{1}{4\pi} \frac{4}{3} \langle
\Gamma | \eth^2 \bar \eth^2 \Gamma \rangle.
\]
One also has that
\[
\langle \Gamma | \eth^2 \bar \eth^2 \Gamma \rangle =\langle \Gamma |
\dot{\Psi } _{0} \delta x \rangle +\langle \Gamma ^{2}
| \dot{\Psi } _{0} \rangle
\]
where we have used that  $\Gamma $
 is orthogonal to the last term appearing 
in the linear nice section equation.

Let us note that from the Schwarz inequality one obtains
\[
| \langle \Gamma | \dot{\Psi } _{0}\delta
x \rangle | \leq || \Gamma || \lambda ||
\delta x ||;
\]
and similarly
\[
\langle \Gamma ^{2} | \dot{\Psi } _{0} \rangle
\leq \lambda || \Gamma|| ^{2}.
\]

From which it is deduced
\[
\sup \left| \Gamma \right| ^{2} \leq \frac{1}{4\pi } \frac{4}{3} \left(
\lambda \left\| \Gamma \right\| ^{2} +\left\| \Gamma \right\| \lambda
\/\left\| \delta x\right\| \right).
\]
It follows from the techniques used in ref. \cite{Moreschi88} and in Section 
V, that  $\Gamma $ is expressible as a series of the form:
\[
\Gamma =\Gamma _{1} \;\lambda +\Gamma _{2} \;\lambda ^{2} + \cdots
\]
which satisfies
\[
\left\| \Gamma _{n} \;\lambda ^{n} \right\| \leq \left( \lambda
\sqrt{\frac{4}{27} } \right) ^{n-1} \;\left\| \Gamma _{1} \;\lambda ^{}
\right\|
\]
with
\[
||\Gamma_1\lambda||\leq \sqrt{\frac{4}{27}} ||\eth^2 \bar \eth^2
\Gamma_1 \lambda||=\sqrt{\frac{4}{27}} ||\dot \Psi_0 \delta x +\Psi_0
+ K^3_1 M_1||\leq \sqrt{\frac{4}{27}} ||\dot \Psi_0 \delta x||\leq
\lambda \sqrt{\frac{4}{27}}||\delta x||
\]
Then the modulus of 
 $\Gamma $
 is bounded by
\[
\left\| \Gamma \right\| \leq \sum\limits_{n=1}^{\infty }\left\| \Gamma
_{n} \;\lambda ^{n} \right\|  \leq \sum\limits_{n=1}^{\infty }\left(
\lambda \sqrt{\frac{4}{27} } \right) ^{n-1} \left\| \Gamma _{1} \;\lambda
^{} \right\| \leq \sum\limits_{n=1}^{\infty }\;\left( \lambda
\sqrt{\frac{4}{27} } \right) ^{n} \left\| \delta x\right\|  
=\frac{\lambda \sqrt{\frac{4}{27} } }{1-\lambda \sqrt{\frac{4}{27} } }
\left\| \delta x\right\|,
\]
for  $\lambda <\sqrt{\frac{27}{4} } \cong 2.598$.

Using this in our main inequality one obtains
\[
\sup \left| \Gamma \right| ^{2} \leq \frac{1}{4\pi } \frac{4}{3}
\sqrt{\frac{27}{4} } \left( \frac{\lambda }{\sqrt{\frac{27}{4} } -\lambda
} \/\right) ^{2} \;\left\| \delta x\right\| ^{2} =\frac{1}{\pi }
\sqrt{\frac{3}{4} } \left( \frac{\lambda }{\sqrt{\frac{27}{4} } -\lambda }
\/\right) ^{2} \;\left\| \delta x\right\| ^{2}.
\]
Taking $\delta x=\delta x_{0} >0$ where  $\delta x_{0} $
 is constant, then one has that
\[
\left\| \delta x\right\| ^{2} =4\;\pi \left( \delta x_{0} \right) ^{2}
\]
and so:
\[
\sup\left| \Gamma \right| ^{2} \leq 2\sqrt{3} \left( \frac{\lambda
}{\sqrt{\frac{27}{4} } -\lambda } \/\right) ^{2} \;\left( \delta x_{0}
\right) ^{2}.
\]

One can see that 
 $\delta g(\delta x_{0} )$
 will be positive if 
\[
\sup \left| \Gamma \right| ^{2} <\left( \delta x_{0} \/\right) ^{2} \;
\]
which is equivalent to
\[
2\sqrt{3} \;\left( \frac{\lambda }{\sqrt{\frac{27}{4} } -\lambda }
\/\right) ^{2} \;<1.
\]

This inequality is satisfied for $\lambda \in (0,\lambda _{*} )$ with
\[ 
\lambda _{*} =\sqrt{\frac{27}{4} } \frac{1}{1+\sqrt{2\sqrt{3} } }
=0.9080.
\]

\emph{Proof 2:}From the Sobolev inequality one obtains
\[
\sup | \Gamma| ^{2} \leq \frac{1}{4\pi } \frac{1}{4} ||\Delta \Gamma || ^{2};
\]
as it can be deduced from the following argument.

Let the regular function  $\Gamma $
 be given by
\[
\Gamma =\sum_{l=2}^{\infty }\sum_{m=-l}^{l}\Gamma ^{lm}
\;Y_{lm} (\zeta ,\bar{\zeta } );
\]
then one has
\begin{align*}
\left| \Gamma (\zeta ,\bar{\zeta } )\right| ^{2}  & =  \left|
\sum\limits_{l=2}^{\infty }\sum\limits_{m=-l}^{l}\ \Gamma ^{lm} \ Y_{lm}
(\zeta ,\bar{\zeta } )  \right| ^{2}  \\
& = \left| \sum\limits_{l=2}^{\infty }\sum\limits_{m=-l}^{l}\ \Gamma
^{lm} \ l(l+1)\frac{1\ }{l(l+1)} Y_{lm} (\zeta ,\bar{\zeta } )  \right|
^{2}  \\
& \leq   \sum_{l=2}^{\infty }\sum_{m=-l}^{l}\ \left|
\Gamma ^{lm} \right| ^{2} \ \left[ l(l+1)\right] ^{2}
\sum_{l=2}^{\infty }\sum_{m=-l}^{l}\ \left[
\frac{1\ }{l(l+1)} \right] ^{2} \left| Y_{lm} (\zeta ,\bar{\zeta }
)\right| ^{2}. 
\end{align*}
Using the inequality deduced from the addition theorem for spherical 
harmonics and that
\[
\left\| \Delta \Gamma \right\| ^{2} =\left[ \sum\limits_{l=2}^{\infty
}\sum\limits_{m=-l}^{l}\;\left| \Gamma ^{lm} \right| ^{2} \;\left[
l(l+1)\right] ^{2}   \right] ;
\]
one deduces
\[
\left| \Gamma \right| ^{2} \leq \frac{1}{4\pi } \left\| \Delta \Gamma
\right\| ^{2} \left[ \sum\limits_{l=2}^{\infty }\;\frac{\;(2l+1)\;}{\left(
l(l+1)\right) ^{2} }  \right] =\frac{1}{4\pi } \;\frac{1}{4} \left\|
\Delta \Gamma \right\| ^{2} .
\]
Then from the linear nice section equation, one obtains
\[
\langle \Gamma |   \eth^2 \bar \eth^2 \Gamma \rangle =\frac{1}{4} \left\| \Delta \Gamma
\right\| ^{2} +\frac{1}{2} \int \nabla _{a} \Gamma \nabla ^{a} \Gamma 
\,\/dS^{2} =\left\langle \Gamma \right.\left| \dot{\Psi } _{0} \;\delta
x\right\rangle +\left\langle \Gamma ^{2} \right.\left| \dot{\Psi } _{0}
\right\rangle
\]
where we have used that  $\Gamma $
 is orthogonal to the last term appearing 
in the linear nice section equation.

Using the inequalities mentioned in Proof 1 and
\[
\left\| \Gamma \right\| ^{2} \leq \left( 2\pi \right) ^{2} \int \nabla
_{a} \Gamma \nabla ^{a} \Gamma  \,\/dS^{2},
\]
coming from the Poincar\'e lemma, one can deduce
\[
4\pi \/\;\sup \left| \Gamma \right| ^{2} +\frac{1}{2\left( 2\pi \right)
^{2} } \left\| \Gamma \right\| ^{2} \leq \lambda \left\| \Gamma \right\|
^{2} +\left\| \Gamma \right\| \/\;\lambda \/\left\| \delta x\right\|;
\]
or equivalently
\[
\sup \left| \Gamma \right| ^{2} \leq \frac{1}{4\pi } \left\{ \left\|
\Gamma \right\| ^{2} \left( \lambda -\frac{1}{8\;\pi ^{2} } \right)
+\left\| \Gamma \right\| \;\lambda \/\left\| \delta x\right\| \right\} .
\]
Substituting the bound for $\left\| \Gamma \right\|$ found above, one obtains
\[
\sup \left| \Gamma \right| ^{2} \leq \frac{1}{4\pi } \left\{ \left(
\frac{\lambda }{\sqrt{\frac{27}{4} } -\lambda } \left\| \delta x\right\|
\right) ^{2} \left( \lambda -\frac{1}{8\;\pi ^{2} } \right) +\left(
\frac{\lambda }{\sqrt{\frac{27}{4} } -\lambda } \left\| \delta x\right\|
\right) \;\lambda \/\left\| \delta x\right\| \right\};
\]
from which it is deduced
\[
\sup \left| \Gamma \right| ^{2} \leq \frac{1}{4\pi } \left(
\frac{\lambda }{\sqrt{\frac{27}{4} } -\lambda } \right) ^{2} \left(
\sqrt{\frac{27}{4} } -\frac{1}{8\;\pi ^{2} } \right) \;\left\| \delta
x\right\| ^{2}.
\]
Taking  $\delta x=\delta x_{0} >0$ where  $\delta x_{0} $
 is constant, one has
\[
\sup \left| \Gamma \right| ^{2} \leq \left( \frac{\lambda
}{\sqrt{\frac{27}{4} } -\lambda } \right) ^{2} \left( \sqrt{\frac{27}{4} }
-\frac{1}{8\;\pi ^{2} } \right) \;\left( \delta x_{0} \right) ^{2} .
\]
Then in order for 
 $\delta g(\delta x_{0} )$ to be positive, one needs
\[
\sup \left| \Gamma \right| ^{2} <\left( \delta x_{0} \/\right) ^{2},
\]
or equivalently
\[
\left( \frac{\lambda }{\sqrt{\frac{27}{4} } -\lambda } \right) ^{2}
\left( \sqrt{\frac{27}{4} } -\frac{1}{8\;\pi ^{2} } \right)<1.
\]
This inequality is satisfied for  $\lambda \in (0,\lambda _{*} )$ with
\[ 
\lambda _{*} =\sqrt{\frac{27}{4} } \left(
\frac{1}{1+\sqrt{\sqrt{\frac{27}{4} } -\frac{1}{8\pi ^{2} } } } \right)
\cong 0.9962.
\]

\emph{Proof 3:} Let the regular function  $\Gamma $
 be given by
\[
\Gamma =\sum\limits_{l=2}^{\infty }\sum\limits_{m=-l}^{l}\;\Gamma ^{lm}
\;Y_{lm} (\zeta ,\bar{\zeta } ).
\]
it was proved in Appendix 2 that
\[
\left| \Gamma (\zeta ,\bar{\zeta } )\right| ^{2} \leq \frac{1}{4\pi }
\;\frac{4}{27} \left\langle D\;\Gamma \right.\left| D\;{}\Gamma
\right\rangle =\frac{1}{4\pi } \;\frac{4}{27} \left\| {}\Gamma \right\|
_{D}^{2}.
\]
If  $\Gamma $ is a solution of the linear nice section equation then
\[
||\Gamma || _{D}^{2} \leq || \dot{\Psi } _{0} (\delta
x+\Gamma )|| ^{2} \leq \lambda ^{2} || \delta x+\Gamma
|| ^{2} \leq \lambda ^{2} || \Gamma || ^{2} +\lambda
^{2} || \delta x|| ^{2};
\]
as it was explained above. Then using the bound
\[
\left\| \Gamma \right\| \leq \frac{\lambda }{\sqrt{\frac{27}{4} }
-\lambda } \left\| \delta x\right\|,
\]
for  $\left( \lambda \sqrt{\frac{4}{27} } \right) <1$
; one obtains
\[
\sup \left| \Gamma \right| ^{2} \leq \frac{1}{4\pi } \frac{4}{27}
\left\{ \lambda ^{2} \;\left( \frac{\lambda }{\sqrt{\frac{27}{4} }
-\lambda } \/\right) ^{2} \;\left\| \delta x\right\| ^{2} +\lambda ^{2}
\;\left\| \delta x\right\| ^{2} \right\}.
\]

Taking  $\delta x=\delta x_{0} >0$ where  $\delta x_{0} $
 is constant, then one has that
\[
\left\| \delta x\right\| ^{2} =4\pi \;\left( \delta x_{0} \right) ^{2}
\]
and so:
\[
\sup \left| \Gamma \right| ^{2} \leq \frac{4}{27} \;\lambda ^{2}
\;\left\{ \left( \frac{\lambda }{\sqrt{\frac{27}{4} } -\lambda } \/\right)
^{2} \;+1\right\} \left( \delta x_{0} \right) ^{2}.
\]

One can see that  $\delta g(\delta x_{0} )$
 will be positive if 
\[
\sup \left| \Gamma \right| ^{2} <\left( \delta x_{0} \/\right) ^{2},
\]
which is equivalent to
\[
\frac{4}{27} \;\lambda ^{2} \;\left\{ \left( \frac{\lambda
}{\sqrt{\frac{27}{4} } -\lambda } \/\right) ^{2} \;+1\right\} <1.
\]
This inequality is satisfied for $\lambda \in (0,\lambda _{*} )$ with
$\lambda _{*} \cong 1.5134$.


\end{document}